\def\be{\begin{equation}}
\def\ee{\end{equation}}
\def\bea{\begin{eqnarray}}
\def\eea{\end{eqnarray}}

\documentclass[aps,unsortedaddress,amsmath,amssymb]{revtex4}
\usepackage{subfigure}
\usepackage{bm}
\usepackage{graphicx}
\begin{document}

\title{Conservation Law for Massive Scale-Invariant Photons in Weyl-Invariant Gravity}
\author{Aradhya Shukla}
\email{ashukla038@gmail.com}
\affiliation{ Indian Institute of Science Education and Research  Kolkata, \\ Mohanpur-741246, India}
\author{Kumar Abhinav}
\email{gokuabhinav@gmail.com}
\affiliation{  S. N. Bose National Centre for Basic Sciences, JD Block, Sector III,\\ Salt Lake, Kolkata-700106, India}
\author{Prasanta K. Panigrahi}
\email{pprasanta@iiserkol.ac.in}
\affiliation{ Indian Institute of Science Education and Research Kolkata,\\ Mohanpur-741246, India}

\vskip 0.5cm
\date{\today} \vskip .8cm
\begin{abstract}
It is demonstrated that a St\"uckelberg-type gauge theory, coupled to the scalar-tensor theory
of gravity, is invariant under both gauge and Weyl transformations. Unlike the pure St\"uckelberg theory,
this coupled Lagrangian has a genuine Weyl symmetry, with a non-vanishing current. The above is true in the
Jordan frame, whereas in the Einstein frame, the same theory manifests as Proca theory in presence of pure
gravity. It is found that broken scale invariance leads to simultaneous spontaneous breaking of the gauge symmetry.
\end{abstract}

\vskip 0.5cm


\maketitle


\section{Introduction}
As is well known, gauge theories display both dynamics and constraints \cite{noether,dirac, sund} with physical quantities being manifestly gauge-invariant. 
In comparison, the Weyl symmetry, although initially introduced as a gauge
symmetry \cite{Weyl1,Weyl2,Weyl3,Weyl4,misc}, does not have the same pivotal role. However, it still retains intriguing features, important 
in the gravity sector. Recently, in the context of the scalar-tensor gravity (STG), Weyl symmetry is 
shown to be a `fake' one, as it arises in the Jordan frame, with vanishing current \cite{JP}. This symmetry gets completely 
eliminated by appropriately dressing the field variables.

In case of gauge theory, massless-ness ensures Weyl invariance. The corresponding current is
conserved, as the trace of the energy-momentum tensor (EMT) vanishes \cite{CCJ}. However,
in the gauge sector, gauge theories expectedly violate Weyl invariance, due to the presence of the mass term.
The generation of mass has itself been  of great interest, with the Higgs mechanism \cite{higgs,higgs1,EB,EB1,GHK}
being the dominant theory for generation of mass for gauge excitations. 
Other approaches for mass generation for gauge particle have been explored, involving the topological terms \cite{LMP,lahiri,HLS,misc1}. 
However, a seemingly massive Proca theory can be imbibed with gauge invariance through suitable redefinition of the field variable, by introducing
a scalar mode \cite{stuckel,glauber,hieg}.

It is important to mention here that the redefinition of metric tensor via a 
frame transformation is quite similar to Stueckelberg's approach for the massive Proca theory,
where the gauge invariance is not manifestly present. Due to this similarity, they are combined 
into a single theory, with combined Weyl and gauge symmetry. 
Gauge fields enter conformal supergravity models through extension of the geometric connection,
leading to combination of Weyl and gauge symmetries. On the other hand, Kaluza-Klein supergravity and string theories,
reducible to conformal gravity, reduce to STG as an effective theory. However, the {\it combination} of STG,
containing non-minimal scalar-gravity coupling term $R\varphi^2$, and Abelian 
gauge field, is a {\it non-supersymmetric} model \cite{P}, whose symmetry properties has not been analyzed in detail before.
We demonstrate that it is possible to attain 
generalized Weyl-invariance for this {\it massive} theory, through suitable coupling with STG, the latter having `fake' Weyl invariance \cite{JP}. Interestingly, in this process, not only the photon is massive, but the STG also acquires non-vanishing conserved generalized Weyl current, owing to its coupling with the gauge sector. Thus the fake Weyl invariance of the STG is changed to a genuine one through interaction with massive gauge field. Earlier, it has been shown that the Weyl invariant STG plays a key role in understanding inflation in the early universe and it yields dimensional gravitational and cosmological constants \cite{kao, HO, IQ, zhang}. Moreover, some inflationary models based on broken scale invariance,
{\it i.e.} the global limit of Weyl symmetry, have also been developed, where the usual scale symmetry of theory, with
suitable choices of variables, manifests as a shift symmetry \cite{cks}.
In purview of this, the present theory may have relevance also in scale-invariant cosmologies.

The paper is organized as follows. In section II, relevant properties of the St\"uckelberg and Proca theory are elucidated. Section II deals with the symmetry properties of STG, wherein the similarity with the 
St\"uckelberg theory is depicted. The combined St\"uckelberg-STG theory is constructed in Section IV, as a
consequence of Weyl-transformed gravitational Proca theory, simultaneously obtaining a massive gauge and genuine Weyl 
invariance.  Further, it is found that when the scale invariance is broken spontaneously, the
gauge symmetry also manifests in the broken symmetry phase.

\section{Massive Vector Field: St\"uckelberg Theory}
The St\"uckelberg Lagrangian density has the gauge invariant kinetic term for the massive vector field, and can be made gauge invariant by coupling  the gauge boson to a scalar field, which transforms linearly. St\"uckelberg mechanism is important because mass generation and gauge symmetry coexist, without taking recourse to Higgs mechanism. Lagrangian density for the Proca theory is given by,
\begin{eqnarray}
{\cal L}_P = -\frac{1}{4}\, F_{\mu\nu} F^{\mu\nu} + \frac{m^2}{2} a_\mu a^\mu,\label{Proca}
\end{eqnarray}
where $F_{\mu\nu} = \partial_\mu a_\nu - \partial_\nu a_\mu$ is the field strength for the vector field $a_\mu$ and $m$ represents its mass. Expectedly, the above Lagrangian density does not respect the usual gauge symmetry, $\delta a_\mu = \partial_\mu\theta$: 
\begin{eqnarray}
\delta \, {\cal L}_P = m^2\, a_\mu\, \partial^\mu\theta.
\end{eqnarray}
The gauge invariance can be restored through  St\"uckelberg's approach, with the new definition of gauge field as,
\begin{eqnarray}
a_\mu \longrightarrow a_\mu = a_\mu + \frac{1}{m}\,\partial_\mu\chi,
\end{eqnarray}
where $\chi$ is the St\"uckelberg scalar field. Substituting the above in Eq. (\ref{Proca}), we get the  St\"uckelberg Lagrangian density:
\bea
{\cal L}_{\rm St} &=& -\frac{1}{4}\, F_{\mu\nu} F^{\mu\nu} + \frac{m^2}{2} a_\mu a^\mu + m a_\mu \partial^\mu\chi 
+ \frac{1}{2} \partial_\mu \chi \partial^\mu \chi \nonumber\\
&=&{\cal L}_p + m a_\mu \partial^\mu\chi 
+ \frac{1}{2} \partial_\mu \chi \partial^\mu \chi,
\eea
which is invariant under the following symmetry transformations: 
\begin{eqnarray}
\delta a_\mu = \partial_\mu \theta, \qquad \quad \delta \chi = - m \theta, \qquad \delta\, {\cal L}_{\rm St} = 0. 
\end{eqnarray}
The N\"other functions for the two fields are,
\bea
&&E_a^\nu=\frac{\partial{\cal L}_{s}}{\partial a_\nu}-\partial_\mu\frac{\partial{\cal L}_{s}}{\partial\left(\partial_\mu a_\nu\right)}\equiv\partial_\mu F^{\mu\nu}+m^2a^\nu+m\partial^\nu\chi,\nonumber\\
&&E_\chi=\frac{\partial{\cal L}_{s}}{\partial\chi}-\partial_\mu\frac{\partial{\cal L}_{s}}{\partial\left(\partial_\mu\chi\right)}\equiv-\partial^2\chi-m\partial_\mu a^\mu,\label{EOM}
\eea 
where the second function can be obtained from the first by taking a four-divergence.

The {\it off-shell} current $X^\mu$ of this theory is obtained, by considering variation
of the Lagrangian under the gauge transformation:
\bea
\delta {\cal L}_{\rm St}&=&\frac{\partial{\cal L}_{s}}{\partial a_\mu}\delta a_\mu+\frac{\partial{\cal L}_{s}}{\partial\left(\partial_\mu a_\nu\right)}\delta\left(\partial_\mu a_\nu\right) + \frac{\partial{\cal L}_{s}}{\partial\chi}\delta\chi+\frac{\partial{\cal L}_{s}}{\partial\left(\partial_\mu\chi\right)}\delta\left(\partial_\mu\chi\right)\nonumber\\
&=&\left(m^2 a^\mu + m\partial^\mu \chi\right)\partial_\mu \theta + \left(-F^{\mu\nu}\right)\partial_\mu\partial_\nu\theta +\left(\partial^\mu\chi+ma^\mu\right)\left(-m\partial_\mu\theta\right)\nonumber\\
&=&0,\label{OfS}
\eea
where, in the second line, we have used expressions given in Eqs. (4) and (5). Thus, $X^\mu=0$.
\paragraph*{}The {\it on-shell} current $K^\mu$ is obtained by using the equations of motion, demanding both 
$E_a^\mu=0=E_\chi$. This changes $\delta {\cal L}_{\rm St}$ to,
\be
\delta_2{\cal L}_{\rm St}=\partial_\mu\left[\frac{\partial{\cal L}_{s}}{\partial\left(\partial_\mu a_\nu\right)}\delta a_\nu
+\frac{\partial{\cal L}_{s}}{\partial\left(\partial_\mu\chi\right)}\delta\chi\right] \equiv \partial_\mu K^\mu,\nonumber
\ee
leading to,
\be
K^\mu= -\left(F^{\mu\nu}\partial_\nu\theta+m\partial^\mu\chi\theta+m^2a^\mu\theta\right).\label{OnS}
\ee
From Eqs. \ref{OfS} and \ref{OnS}, N\"other's current for the St\"uckelberg system is found to be,
\be
J^\mu:=X^\mu-K^\mu=F^{\mu\nu}\partial_\nu\theta+m\partial^\mu\chi\theta+m^2a^\mu\theta,\label{NS}
\ee
satisfying the conservation law,
\bea
\partial_\mu J^\mu&=&\left(\partial_\mu F^{\mu\nu}+m^2a^\nu+m\partial^\nu\chi\right)\partial_\nu\theta +\left(\partial^2\chi+m\partial_\mu a^\mu\right)m\theta\nonumber\\
&=&E_a^\mu\delta a_\mu+\left(-E_\chi\right)(-\delta\chi)\equiv0,\label{Cons}
\eea
where, Eq. (\ref{EOM}) has been applied at the end.

\subsection{Finite gauge transformation in St\"uckelberg's approach} 
Invariance of the Lagrangian density for the St\"uckelberg theory is independent of the gauge parameter being finite 
or infinitesimally small. As St\"uckelberg theory is obtained from the Proca theory through the {\it finite} gauge transformation,
\be
a_\mu\rightarrow a_\mu-\frac{i}{g}V^{-1}\partial_\mu V,\quad V=\exp\left(i\frac{g}{m}\chi\right),\label{FGT1} 
\ee 
which takes ${\cal L}_P$ to ${\cal L}_{\rm St}$. The next set of gauge transformations,
\bea
a_\mu \rightarrow a_\mu+\frac{i}{h}U^{-1}\partial_\mu U,\qquad \quad
V \rightarrow U V,\quad U=\exp\left(-i\frac{h}{m}\theta\right),\label{FGT2}
\eea
yields,
\be
{\cal L}_{\rm St}\rightarrow {\cal L}_{\rm St}+\left(1-\frac{h}{g}\right)\left(ma_\mu+\partial_\mu\chi\right)\partial^\mu\theta + 2\left(1-\frac{h}{g}\right)^2\partial_\mu\theta\partial^\mu\theta.\label{StF}
\ee
which defines a symmetry under the parametric condition $h=g$. 
\paragraph*{}The same property, namely,
\be
{\cal L}_P\rightarrow{\cal L}_{\rm St}\rightarrow{\cal L}_{\rm St},\label{TransG}
\ee
for successive gauge transformations, with respect to both {\it infinitesimal} parameters 
$\chi$ and $\theta$ respectively, where ${\cal L}_{\rm St}$ is defined {\it without} the
quadratic term in $\chi$, prevails. Therefore, Eq. (\ref{TransG}) holds for both finite and 
infinitesimal gauge transformations.

\subsection{Energy-momentum tensor for the St\"uckelberg theory}
The symmetric energy-momentum (EM) tensor, including the Belinfant\'e term, is,
\be
{\cal T}^{\mu\nu}=2\frac{\delta{\cal L}}{\delta g_{\mu\nu}}-g^{\mu\nu}{\cal L}.\label{EM1}
\ee
For the St\"uckelberg theory, the same is found to be,
\be
{\cal T}^{\mu\nu}_{\rm St}= -F^{\mu\alpha}F^\nu_{~\alpha}+m^2a^\mu a^\nu+\partial^\mu\chi\partial^\nu\chi + m\left(a^\mu\partial^\nu\chi+a^\nu\partial^\mu\chi\right)-g^{\mu\nu}{\cal L}_{\rm St},\label{EMSt}
\ee
with the corresponding trace,
\be
g_{\mu\nu}{\cal T}^{\mu\nu}_{\rm St}= -\left(m^2a_\mu a^\mu+\partial^\mu\chi\partial_\mu\chi+2ma^\mu\partial_\mu\chi\right).\label{Tr1}
\ee
Clearly, the theory is not Weyl-invariant, expectedly for being massive, but also for containing
the scalar field $\chi$. On adding a superpotential term \cite{Deser, CCJ}, the improved EM
tensor becomes,
\bea
\Theta^{\mu\nu}_{\rm St}={\cal T}^{\mu\nu}_{\rm St}+\frac{1}{6}\left(g^{\mu\nu}\partial_\alpha\partial^\alpha
-\partial^\mu\partial^\nu\right)\chi^2,\label{Th1}
\eea
leading to the trace,
\be
g_{\mu\nu}\Theta^{\mu\nu}_{\rm St}=-\left(m^2a_\mu a^\mu+2ma^\mu\partial_\mu\chi\right),\label{Tr2}
\ee
after using the equation of motion for the scalar field, in the Lorentz gauge. Thus, the on-shell trace
of the St\"uckelberg EM tensor vanishes for $m=0$, as required.

\section{Emergence of Scalar-Tensor Theory from Pure Gravity Theory}
Let us now consider the (3 + 1)-dimensional Einstein-Hilbert (EH) action,
\bea
S =  - \int d^4x\,{\cal L}_{0} = - \int d^4x\, \frac{1}{12k}\sqrt{-g}\,R,\label{EH}
\eea
where $k = (16\pi G)$ is an overall constant and $R$ is the Ricci scalar defined as $R = g^{\mu\nu}\, R_{\mu\nu}$, with $g^{\mu\nu}$ and $R_{\mu\nu}$
 being the metric and Riemann curvature tensor, respectively. It is important to point that the curvature present in the free gravity theory is due 
to the metric tensor and mass of the background, because of which the above Lagrangian density changes under the Weyl scale transformation: $g^{\mu\nu}\rightarrow e^{2\theta} g^{\mu\nu}$, where $\theta$ is a local parameter. For $\theta$ being infinitesimally small, the Weyl 
symmetry becomes $\delta g^{\mu\nu} = 2\theta g^{\mu\nu}$, under which the Lagrangian density transforms as,
\be
\delta {\cal L}_0 = \frac{2}{k}\theta \sqrt{-g} R.
\ee 
The symmetry can be restored by redefining the metric tensor as $g_{\mu\nu} = \varphi^2 g_{\mu\nu}$, where $\varphi$ is a scalar field. This redefinition is similar to the St\"uckelberg's approach, discussed in the previous section. It is important to point out that to maintain the Weyl invariance, the metric tensor has been scaled by a local scalar field, whereas in Proca theory, a derivative of scalar field has been added to redefine the vector field, for the desired gauge invariance.

Under the locally scaled metric tensor, the field variables present in the theory change accordingly as,
\bea
g_{\mu\nu} \rightarrow  \varphi^2\, g_{\mu\nu}, \qquad g^{\mu\nu}  \rightarrow \varphi^{-2}\, g^{\mu\nu},\qquad
 \sqrt {-g} \longrightarrow  \varphi^4\sqrt{-g},
\eea
with the modified Ricci scalar \cite{Deser},
\be
\sqrt {-g}R \longrightarrow \sqrt {-g} R \,\varphi^2 + 6\sqrt {-g} g^{\mu\nu}\,\partial_\mu \varphi \partial_\nu \varphi.\label{23}
\ee 
Finally, substituting the changes in field variables, we get a modified Lagrangian, known as the scalar-tensor Lagrangian density: 
\be
{\cal L}_{\rm ST}=\frac{1}{\kappa} \Big[\frac{1}{12}\sqrt{-g}R\varphi^2+\frac{1}{2}\sqrt{-g}g^{\mu\nu}\partial_\mu\varphi\partial_\nu\varphi \Big].\label{JP}
\ee
The infinitesimal Weyl symmetry,
\be
\delta g^{\mu\nu} = 2\theta\, g^{\mu\nu}, \qquad \delta\varphi = \theta\, \varphi,
\ee
changes the Lagrangian by a total derivative, leading to  {\it off-shell} contribution
to the conserved current as,
\be
X^\mu=\frac{1}{2 \kappa}\sqrt{-g}\varphi^2g^{\mu\nu}\partial_\nu\theta.\label{26}
\ee
The use of Euler-Lagrange equations for $\varphi$ and $g_{\mu\nu}$, re-casts the variation of the Lagrangian as a
total derivative, leading to the {\it on-shell} contribution to the Weyl current, 
\be
K^\mu=\frac{1}{2 \kappa}\sqrt{-g}\varphi^2g^{\mu\nu}\partial_\nu\theta.
\ee
As both these contributions are equal, the conventional conserved N\"other Weyl current vanishes \cite{JP}: 
\bea
J^\mu =K^\mu - X^\mu = 0,
\eea
deeming the Weyl symmetry as a fake one. The situation is unchanged \cite{JP} upon application of N\"other's
second theorem \cite{noether, Weyl1, Weyl2, Weyl3, Weyl4}, appropriate for local symmetries, such as the Weyl symmetry here.
The present aim is to obtain an extended theory having genuine generalized symmetry, instead of the `fake' one, thereby obtaining a non-vanishing current. Although the matter of extended Weyl symmetries have been discussed earlier, with additional fields, having general coupling to STG \cite{New1} and massive excitations \cite{New2}, the issue of the conserved current was not addressed. In section IV, we construct the simplest example of a theory, non-trivially coupled to the STG, yielding an extended, but genuine, Weyl symmetry.

\subsection{Finite Weyl transformations in scalar-tensor theory}
The scalar-tensor Lagrangian density (${\cal L}_{\rm ST}$) is obtained from the free gravity
Lagrangian density (${\cal L}_{0}$) through a finite Weyl transformation, given in Eq. (20).
Let us consider the generic finite {\it local} scaling of the form,
\be
g^{\mu\nu}\rightarrow \psi^{-n} g^{\mu\nu},\quad g_{\mu\nu}\rightarrow \psi^n g_{\mu\nu}\quad\varphi\rightarrow\psi\varphi,\label{N01}
\ee
with $\psi$ being local and finite and $n$ being the numerical power. Under such scaling, the terms of
${\cal L}_{\rm ST}$ transform as,
\bea
&&\sqrt{-g}\frac{1}{12\kappa} R( g) \varphi^2 \rightarrow \sqrt{-g}\frac{1}{12\kappa} \Big[ \psi^{n+2}\varphi^2R(g)
+\frac{3}{2}n(n+4) \psi^n \varphi^2 g^{\mu\nu} \partial_\mu \psi \partial_\nu \psi\nonumber\\
&&\qquad\qquad\qquad\qquad
+ 6n g^{\mu\nu} \psi^{n+1} \varphi \partial_\mu \psi \partial_\nu \varphi \Big]- \frac{n}{4\kappa}\partial_\mu \Big[ \sqrt{-g} \varphi^2 \psi^{n+1} \partial^\mu \psi \Big],  \nonumber\\
&&\sqrt{- g} \frac{1}{2\kappa} g^{\mu\nu}\partial_\mu\varphi \partial_\nu \varphi 
\rightarrow \sqrt{-g} \frac{1}{2\kappa}\psi^n g^{\mu\nu} \Big[ \varphi^2 \partial_\mu \psi \partial_\nu \psi
+ 2 \psi \varphi \partial_\mu \psi \partial_\nu \varphi \nonumber\\
&&\qquad\qquad\qquad\qquad+ \psi^2 \partial_\mu \varphi \partial_\nu \varphi \Big].\label{N02}
\eea
For the present case,
\be
g^{\mu\nu}\rightarrow\psi^2g^{\mu\nu},\quad g_{\mu\nu}\rightarrow\psi^{-2}g_{\mu\nu},\quad\varphi\rightarrow\psi\varphi,
\ee
corresponding to $n=-2$, the general expressions reduce to,
\bea
&&\sqrt{-g}\frac{1}{12\kappa}R\varphi^2 \rightarrow \sqrt{-g}\Big[\frac{1}{12\kappa}R\varphi^2 - \frac{1}{2\kappa} \psi^{-2} \varphi^2 g^{\mu\nu} 
\partial_\mu \psi \partial_\nu \psi- \frac{1}{\kappa} g^{\mu\nu} \psi^{-1} \varphi \partial_\mu \psi \partial_\nu \varphi \Big]\nonumber\\
&&\qquad~~~~~~~~~~~~~~+ \partial_\mu \Big[ \sqrt{-g} \frac{1}{2\kappa} \frac{\varphi^2} {\psi} g^{\mu\nu}\partial_\nu \psi \Big], \nonumber\\
&&\sqrt{-g} \frac{1}{2\kappa}g^{\mu\nu}\partial_\mu\varphi\partial_\nu\varphi \rightarrow \sqrt{-g}\Big[ \frac{1}{2\kappa}g^{\mu\nu}\psi^{-2} \varphi^2 \partial_\mu \psi\partial_\nu\psi 
+ \frac{1}{\kappa} g^{\mu\nu} \psi^{-1} \varphi \partial_\mu \psi \partial_\nu \varphi\nonumber\\
&&\qquad~~~~~~~~~~~~~~~~~~~~~+ \frac{1}{2\kappa} g^{\mu\nu}\partial_\mu \varphi \partial_\nu \varphi \Big].\label{FCT1}
\eea
On combining both parts, we have,
\be
{\cal L}_{\rm ST}\rightarrow{\cal L}_{\rm ST} 
+ \partial_\mu \Big[ \sqrt{-g} \frac{1}{2\kappa} \frac{\varphi^2} {\psi} g^{\mu\nu}\partial_\nu \psi \Big].\label{FCT2}
\ee

Therefore, ${\cal L}_{\rm ST}$ changes only by a total derivative under Weyl scaling by a finite local function $\psi(x)$,
leaving the corresponding action unchanged as expected. Therefore, the STG is invariant under both finite and 
infinitesimal Weyl transformations, just like the gauge transformation of St\"uckelberg theory. This intuitively enables 
us to combine both the theories through identification of the corresponding local transformation parameters. As a check, for
infinitesimal Weyl transformation, $\psi=\exp(\theta), \theta\ll 1$, one has,

$${\cal L}_{\rm ST}\rightarrow{\cal L}_{\rm ST}+ \partial_\mu \Big[ \sqrt{-g} \frac{1}{2\kappa} \varphi^2 g^{\mu\nu}\partial_\nu \theta \Big],$$
yielding the expression of the off-shell contribution for the current given in Eq. (\ref{26}) for infinitesimal Weyl
scaling.

\subsection{Energy-momentum tensor for scalar-tensor theory} 
In this subsection, we derive the EMT for the STG and show that it is automatically symmetric, as well as
traceless in nature. The EMT for a massless scalar field is given by,
\be
{\cal T}^\varphi_{\mu\nu}\equiv\partial_\mu\varphi\partial_\nu\varphi 
- g_{\mu\nu}\frac{1}{2}g^{\alpha\beta}\partial_\alpha\varphi\partial_\beta\varphi,\label{EMTS}
\ee
which is symmetric but not traceless. It was postulated that introduction of the additional {\it transverse} 
part, $\frac{1}{6}\left(g^{\mu\nu}\nabla^\rho\nabla_\rho-\nabla^\mu\nabla^\nu\right)\varphi^2$ makes the 
EMT traceless, while preserving the original Poincar\'e generators of the theory, thereby keeping the dynamical
observables unchanged \cite{CCJ}. The variation of the gravitational part of the theory,
with respect to the metric, leads to the corresponding contribution to the EMT as,
\bea
&&{\cal T}_g^{\mu\nu}=G^{\mu\nu}\varphi^2+\left(g^{\mu\nu}\nabla^\rho\nabla_\rho-\nabla^\mu\nabla^\nu\right)\varphi^2;\label{EMTG}\\
&&G_{\mu\nu}=R_{\mu\nu}-\frac{1}{2}g_{\mu\nu}R.\nonumber
\eea
Here, $\nabla^\mu$ is the covariant derivative with respect to the gravitational metric. This yields to the complete EMT of the theory,
\bea
\theta^{\mu\nu}&=&{\cal T}_\varphi^{\mu\nu}+{\cal T}_g^{\mu\nu}\nonumber\\
&\equiv&\partial^\mu\varphi\partial^\nu\varphi-g^{\mu\nu}\frac{1}{2}g^{\alpha\beta}\partial_\alpha\varphi\partial_\beta\varphi+\frac{1}{6}G^{\mu\nu}\varphi^2\nonumber\\
&+&\frac{1}{6}\left(g^{\mu\nu}\nabla^\rho\nabla_\rho-\nabla^\mu\nabla^\nu\right)\varphi^2.\label{EMTJP}
\eea
Therefore, the `transverse' extension to the scalar part, required for Weyl symmetry, is 
automatically provided by the gravitational part. Hence, the complete scalar-tensor theory is Weyl-invariant.

\section{A Combined Theory}
Motivated by the similarity between the roles of Weyl transformation in case 
of gravity and that of gauge transformation in case of massive photon, it is tempting to hope
for a larger picture, which can accommodate these two sectors. This has indeed been found to be 
possible. Starting in the Einstein frame, let us define the action,
\be
{\cal S}_{\rm EHP}=-\int d^4x\sqrt{-g}\Bigl[\frac{1}{12\kappa}R-\frac{1}{4}F_{\mu\alpha}F_{\nu\beta}g^{\mu\nu}g^{\alpha\beta}
+\frac{m^2}{2}a_\mu a_\nu g^{\mu\nu}\Bigr].\label{EHP1}
\ee
This keeps the STG action dimensionless, as required, provided we consider the overall constant
$(16\pi G)^{-1}$ of dimension $[m]^2$, with $G$ being the gravitational constant, of $[m]^2$.
The corresponding action, though dimensionally admissible, is neither Weyl nor gauge invariant. More specifically,
the first and third terms change under Weyl scaling, whereas the third changes under gauge transformation too.
The second term is gauge invariant, and also Weyl invariant, as we are in (3+1)-dimensions. The overall Weyl non-invariance
is further reflected by the non-zero trace of the complete EM tensor,
\be
{\cal T}^{\mu\nu}=\frac{1}{6\kappa}G^{\mu\nu}-F^{\mu\alpha}F^\mu_{~\alpha}+m^2a^\mu a^\nu-g^{\mu\nu}{\cal L}_{\rm St},\label{EMTN1}
\ee
leading to,
\be
{\cal T}^\mu_{~\mu}=-\frac{1}{6\kappa}R-m^2a^\mu a_\mu.\label{EMTN2}
\ee
A simultaneous Weyl and gauge transformation,
\be
g_{\mu\nu}\rightarrow\varphi^2g_{\mu\nu},\quad a_\mu\rightarrow a_\mu-\partial_\mu\log\varphi,\label{GCT1}
\ee
leads to the following Lagrangian,
\bea
{\cal L}_{\rm STSt}&=& \sqrt{-g}\Bigl[\frac{1}{12\kappa}R\varphi^2+\frac{1}{2\kappa}\partial_\mu\varphi\partial_\nu\varphi g^{\mu\nu}
-\frac{1}{4}F_{\mu\alpha}F_{\nu\beta}g^{\mu\nu}g^{\alpha\beta}\nonumber\\
&+& m^2\varphi^2\Big(\frac{1}{2}a_\mu a_\nu+\frac{1}{2}\partial_\mu\log\varphi\partial_\nu\log\varphi  
- a_\mu\partial_\nu\log\varphi\Big)g^{\mu\nu}\Bigr],\label{JPSt1}
\eea
in the {\it Jordan} frame. The above action can effectively be viewed as the sum of STG and
St\"uckelberg actions, with a suitable coupling modifying the gauge mass term, including the
field $\varphi$. In that sense, it is different from St\"uckelberg theory. The
above action is invariant under the set of `combined' transformations,
\be
g^{\mu\nu}\rightarrow\psi^2g^{\mu\nu},\quad\varphi\rightarrow\psi\varphi,\quad a_\mu\rightarrow a_\mu+\partial_\mu\log\psi,\label{GCT2}
\ee
with $\log\psi$ being small,
which is clear from the treatment of the previous sections. The combined transformations are defined in terms
of a {\it single} local parameter ($\varphi$ or $\psi$), with Weyl and gauge subsets being independent, as the
gauge field is a Weyl scalar. Further, ${\cal L}_{\rm STSt}$ goes back to ${\cal L}_{\rm EHP}$ for
$\psi=\varphi$.

The Lagrangian in Eq. (\ref{JPSt1}) can be re-expressed as,
\bea
{\cal L}_{\rm STSt}=\sqrt{-g}\Bigl[\frac{1}{12\kappa}R\varphi^2+\frac{1}{2\kappa}\partial_\mu\varphi\partial_\nu\varphi g^{\mu\nu}
-\frac{1}{4}F_{\mu\alpha}F_{\nu\beta}g^{\mu\nu}g^{\alpha\beta}
+\frac{m^2}{2}g^{\mu\nu}{\cal D}_\mu\varphi{\cal D}_\nu\varphi\Bigr],\label{JPSt2}
\eea
with ${\cal D}_\mu=\partial_\mu-a_\mu$ being the covariant derivative corresponding to the well-known
R-symmetry (or $U(1)_R$ symmetry), analogous to that of the standard $U(1)$
gauge theory. The last term above is invariant under the combined transformation, with the change in the
term $\sqrt{-g}g^{\mu\nu}$ compensating for the same in ${\cal D}_\mu\varphi{\cal D}_\nu\varphi$. This is
unlike the scalar QED, where complex-conjugation in $\left(D_\mu\phi\right)^\dagger D^\mu\phi$ maintains
its invariance, with $D_\mu=\partial_\mu-ia_\mu$ and complex scalar field $\phi$. The $U(1)$ coupling of
gauge field with complex scalar field $\phi$ physically represents the interaction of particle-antiparticle
through exchange of photon. In the present case, as the gauge transformation is aided by the Weyl
transformation to restore the overall symmetry, the real scalar field interacts, {\it without} any
charge, via $U(1)_R$ photon exchange, through its coupling with the metric (gravity). Such gauge interactions 
are common in supergravity models \cite{N001}. Therefore,
from the interaction point-of-view too, the Weyl+gauge transformation corresponds to the complete
symmetry of the theory. It is worthwhile to mention that the relation between the Weyl symmetry and
the gauge symmetry is similar to the color-flavor locking in quantum chromodynamics (QCD) \cite{wil}.

\subsection{Equations of motion}
The combined action in Eq. (\ref{JPSt2}) yields the respective equations of motion
for the gauge ($a_\mu$), scalar ($\varphi$) and gravitational ($g_{\mu\nu}$) fields as,
\bea
&&\nabla_\mu F^{\mu\nu}=-m^2\varphi^2a^\nu+m^2\varphi\partial^\nu\varphi,\nonumber\\
&&\Delta\varphi=\frac{1}{1+\kappa m^2}\left[\frac{1}{6}R+\kappa m^2\left(\nabla_\mu a_\nu+a_\mu a_\nu\right)g^{\mu\nu}\right]\varphi,\nonumber\\
&&{\cal T}_{\mu\nu}=\frac{1}{\kappa}\Big[\frac{1}{12}G_{\mu\nu}\varphi^2+\frac{1}{2}\partial_\mu\varphi\partial_\nu\varphi-\frac{1}{4}g_{\mu\nu}g^{\alpha\beta}\partial_\alpha\varphi\partial_\beta\varphi+\frac{1}{12}\left(g_{\mu\nu}\Delta-\nabla_\mu\nabla_\nu\right)\varphi^2\Big]\nonumber\\
&&\qquad\quad\quad-\frac{1}{2}F_\mu^{~\alpha}F_{\nu\alpha}+\frac{m^2}{2}\varphi^2\bar{a}_\mu\bar{a}_\nu-\frac{1}{4}g_{\mu\nu}\left(-\frac{1}{2}F_{\alpha\beta}F^{\alpha\beta}+m^2\varphi^2\bar{a}_\alpha\bar{a}^\alpha\right)\nonumber\\
&&\qquad\equiv0;\label{EOM02}
\eea
with,
\bea
&&\Delta\varphi=\frac{1}{\sqrt{-g}}\partial_\mu\left[\sqrt{-g}g^{\mu\nu}\partial_\nu\varphi\right],\qquad \bar{a}_\mu
=a_\mu-\partial_\mu\log\varphi.\label{Notation1}
\eea
The trace of the {\it off-shell} total energy-momentum tensor ${\cal T}^\mu_{~\mu}$, by utilizing equation of
motion for the scalar field, turns out as,
\be
{\cal T}^\mu_\mu=\frac{m^2}{2\left(1+\kappa m^2\right)}\varphi^2\left[g^{\mu\nu}\nabla_\mu a_\nu+g^{\mu\nu}a_\mu a_\nu
-\frac{1}{6}R\right]-\frac{m^2}{2}g^{\mu\nu}{\cal D}_\mu\varphi{\cal D}_\nu\varphi.\label{GF01}
\ee
Thus, the presence of photon mass $m$ [Eq. (\ref{EHP1})] breaks the naive Weyl invariance, as
physically expected. The Lagrangian in Eq. (\ref{JPSt2}) is further non-invariant under pure gauge 
transformation, in absence of scaling of the metric. The individual violations of both Weyl
and gauge symmetry owe to the last term in Eq. (\ref{JPSt2}), which corresponds to the Proca
mass term, in addition to being coupled with $\varphi^2$. This term is invariant only under 
combined Weyl-gauge transformation of Eq. (\ref{GCT2}), and so is the full theory. Therefore,
massless electrodynamics and gravity can co-exist with independent Weyl and gauge symmetries
(in the Jordan frame). Introduction of photon mass, though breaks {\it both} of these symmetries
individually, it re-adjusts the system to be invariant under an extended symmetry, which is
the combined Weyl-gauge symmetry.

The crucial role of the mass term to extend the symmetry, unique from the
naive sum of Weyl and gauge symmetries, can be clearly understood 
from the equation of motion for the scalar field. The RHS represents a mass term
contributed by gravity, in violation to the equivalence principle. Further, it includes
contributions due to gauge coupling. Therefore, the theory is symmetric {\it only} under the complete
Weyl-gague transformation, with tensor, vector and scalar parts compensating for each-other. In case of the scalar field
itself being massive, it is known to induce a contribution equivalent to the cosmological constant \cite{CCJ}. For the
present mass-less scalar field, no such shift occurs. Instead, it is the coupling to the gauge field that provides
a mass-like contribution. The same is further reflected in the non-vanishing RHS of Eq. (\ref{GF01}), making the complete
EMT of the theory trace-full. However, this contribution entirely comes from the gauge-coupling, as represented by the 
overall multiplication by the coupling strength $m^2$. Indeed, for `pure' Weyl invariance, the EMT must be trace-less 
{\it by construction}, {\it i.e.} off-shell in the gravitational sector. Therefore, although the gravitational EOM in Eqs. 
\ref{EOM02} yields ${\cal T}_{\mu\nu}=0$, and thereby ${\cal T}^\mu_\mu=0$, the theory is {\it not} Weyl-invariant. This
is true for any theory coupled to gravity, as the gravitational EOM always results in vanishing of total EMT. Applying
all three EOMs, an additional condition,

$${\bar D}^2\varphi^2\equiv\varphi^2 g^{\mu\nu}\left(\nabla_\mu a_\nu+a_\mu a_\nu\right),\qquad{\bar D}_\mu=\nabla_\mu-a_\mu,$$
is obtained, merely stating that all fields are dynamically {\it not} independent \cite{Own}.

Intuitively, coupling with gauge field can make the scalar field `massive', but that mass
will be gauge-dependent in general, as is well-known from the self-energy corrections in quantum electrodynamics (QED) \cite{IZ}.
The non-minimal gravitational coupling is known to restore the Weyl invariance of the mass-less scalar field in STG.
In the present case, due to gauge coupling, the scalar field acquires a mass-like term that breaks the naive
Weyl invariance. But the additional St\"uckelberg-like gauge structure 
{\it extends} the same to the unique Weyl-gauge symmetry. The key to this `combined' symmetry 
is clearly the identification of $\varphi$ in both the Weyl and gauge sectors. The analysis of the present theory,
with an appropriate gauge-fixing, is under investigation and will be reported elsewhere.  
\paragraph*{}For the pure STG [Eq. (\ref{JP})], the transverse contribution of the $\sqrt{-g}R\varphi^2$ to the
energy-momentum tensor restores Weyl invariance of the dynamic scalar part. Therefore, the question arises regarding what
restores the Weyl invariance of the additional dynamic scalar contribution,

$$\sqrt{-g}\varphi^2g^{\mu\nu}\partial_\mu\log\varphi\partial_\nu\log\varphi =\sqrt{-g}g^{\mu\nu}\partial_\mu\varphi\partial_\nu\varphi,$$
coming from the St\"uckelberg contribution. In reality, this part is made Weyl invariant by the same transverse
contribution, as the Weyl invariance restoration of the dynamic scalar part is {\it on-shell}, {\it i.e.}, 
the equation of motion for the scalar field has to be used. In the combined theory the scalar field
equation of motion gets contribution from {\it both} the dynamic parts, as seen in Eq. (\ref{EOM02}). 
On utilization of the same, the transverse part of the energy-momentum tensor exactly compensates
for the Weyl non-invariance of both dynamic scalar parts.

\subsection{Spontaneously broken scale symmetry}
For the possibility of spontaneous symmetry breaking of the scale invariance, we now consider the global limit of Weyl symmetry 
(scale symmetry), after adding a potential term $V(\varphi) = \beta\varphi^4$, having a non-vanishing vacuum expectation value (VEV): 
$\langle \varphi \rangle^2 =  \Omega^2/2\gamma$, for a generic Lagrangian density,
\begin{eqnarray}
{\cal L}_{\rm STSt1}=&&\sqrt{-g}\Big[\frac{\gamma}{\kappa}R\varphi^2+\frac{1}{2\kappa}\partial_\mu\varphi\partial_\nu\varphi g^{\mu\nu} + V(\varphi)
-\frac{1}{4}F_{\mu\alpha}F_{\nu\beta}g^{\mu\nu}g^{\alpha\beta}\nonumber\\
&&\qquad+ m^2\varphi^2\Bigl(\frac{1}{2}a_\mu a_\nu +\frac{1}{2}\partial_\mu\log\varphi\partial_\nu\log\varphi
-a_\mu\partial_\nu\log\varphi\Bigr)g^{\mu\nu}\Big],\label{JPSt3}
\end{eqnarray}
where $\gamma$ is a dimensionless parameter. Although we are presently interested with global scale symmetry, the
potential term $\sqrt{-g}V(\varphi)$ is invariant also under Weyl scaling which is local, and known to effectively represent
the cosmological constant in STG models \cite{CCJ}. By suitably re-scaling the metric and gauge fields:
\begin{eqnarray}
g_{\mu\nu} \rightarrow \frac{\Omega^2}{2\gamma} \varphi^{-2} g_{\mu\nu}, \qquad a_\mu \rightarrow a_\mu + \partial_\mu \log \varphi,
\end{eqnarray}
one gets back the Lagrangian density in Einstein frame with some modifications as,
\begin{eqnarray}
{\cal L}_{\rm EHP1} = \sqrt{-g}\Big[\frac{\Omega^2}{2\kappa} R + \frac{1}{2}g^{\mu\nu} \partial_\mu \xi \partial_\mu \xi + V(\xi)- \frac{1}{4}F_{\mu\alpha}F_{\nu\beta}g^{\mu\nu}g^{\alpha\beta} + \frac{ \Omega^2}{2 \gamma}\Big(\frac{m^2}{2}a_\mu a_\nu \Big) g^{\mu\nu} \Big],\label{51}
\end{eqnarray}
where $V(\xi) = \frac{1}{4}\gamma^{-2}\Omega^4\varphi(\xi)^{-4}V[\varphi(\xi)]$. The relation between the original scalar ($\varphi$) and 
Einstein frame scalar ($\xi$) is,
\begin{eqnarray}
\varphi (\xi) = \langle \varphi\rangle\, \exp\Big( \frac{\sqrt {\tilde \gamma}\xi}{ \Omega} \Big), \qquad \frac{1}{\tilde\gamma} = \frac{1}{2\gamma} - 6
\end{eqnarray} 
with boundary condition $\varphi (\xi = 0) = \langle \varphi \rangle$. In the Einstein frame, the original scale symmetry becomes the shift symmetry,
\begin{eqnarray}
\xi \longrightarrow \bar \xi = \xi +  \frac{\Omega}{\sqrt {\tilde \gamma}}\, \theta, \qquad \quad a_\mu = a_\mu.
\end{eqnarray}
Here, $\theta$ is presently a global parameter, corresponding to the scale symmetry, instead of the Weyl symmetry, which is local.
Under the shift symmetry, Lagrangian density in Eq. (\ref{51}) remains invariant. Hence, ${\cal L}_{\rm EHP1}$ is the Lagrangian for the spontaneously broken
scale symmetry, where the above shift symmetry is the part of the original scale invariance and $\varphi$ is the Goldstone boson for the broken scale invariance. The above analysis is valid for in generic $n$-dimensions, unlike the combined Weyl-gauge symmetry before, which require $n = (3+1)$-dimensions.

\subsection{N\"other's conserved current}
As the N\"other current corresponds to continuous symmetries, allowing for infinitesimal transformations,
one can re-parameterize the parameter for the {\it second} Weyl transformation as,

$$\psi=\exp\left(\frac{\theta}{m}\right).$$
A similar parameterization leads to,

$$\varphi=\exp\left(-\frac{\chi}{m}\right).$$
These re-definitions are adopted, as for a dimensionless Weyl scalar field, representing pure scaling of the metric, 
$\theta$ and $\chi$ must have dimension of mass ($[m]^1$), a fact essential for identifying
$\chi$ as the physical Nambu-Goldstone field of the St\"uckelberg theory \cite{stuckel}.
Then, the infinitesimal Weyl-gauge variations become,
\be
\delta g^{\mu\nu} = 2\frac{\theta}{m} g^{\mu\nu}, \qquad 
\delta a_\mu =  \frac{1}{m}\partial_\mu\theta,\qquad \delta\chi = -\theta.\label{GTN03}
\ee
As in section II, the off-shell and on-shell variations, respectively, lead to,
\bea
\delta_1{\cal L}_{\rm STSt}=&&\partial_\mu\left(\frac{1}{\kappa}X^\mu\right),\quad X^\mu=\frac{1}{2m}\sqrt{-g}\varphi^2g^{\mu\nu}\partial_\nu\theta,\nonumber\\
\delta_2{\cal L}_{\rm STSt}=&&-\partial_\mu\Big[\sqrt{-g}\Big\{\frac{1}{m}F^{\mu\nu}\partial_\nu\theta+\exp\left(-2\frac{\chi}{m}\right)\theta\left(ma^\mu+\partial^\mu\chi\right)\Big\}\Big] + \partial_\mu\left(\frac{1}{\kappa}X^\mu\right);\label{VarN01}
\eea
yielding the conserved N\"other current,

\bea
&& \delta_1{\cal L}_{\rm STSt}-\delta_2{\cal L}_{\rm STSt}=\partial^\mu J_\mu=0,\nonumber\\
&&J_\mu = \sqrt{-g}\Big\{\frac{1}{m}F_{\mu\alpha}g^{\alpha\beta}\partial_\beta\theta+\exp\left(-2\frac{\chi}{m}\right)m\theta\Big(a_\mu+\frac{1}{m}\partial_\mu\chi\Big)\Big\}.\label{NC01}
\eea
having a non-vanishing expression.

\subsection{A parametric duality}
The crucial feature of being able to construct the combined theory in Eq. (\ref{JPSt2}) is 
the Weyl invariance of the mass term. It demands only that $\sqrt{-g}\varphi^2$ is
dimensionless and $\partial_\mu\varphi$ is of dimension one ($[m]^1$). This, however, leaves
a freedom of choice for dimensions for both $\sqrt{-g}$ and $\varphi$, expected from
Weyl invariance. On the other hand, dimension of $\partial_\mu\varphi$ remains unaffected 
by any such choice, which is essential for the gauge part of the combined transformations.

From the Weyl transformation $g_{\mu\nu}\rightarrow\varphi^2g_{\mu\nu}$, 
a particular choice can be of $\varphi$ being a $[m]^1$ real scalar field, thereby representing
physical dynamics. As a consequence, it is required that,
\be
ds^2=g_{\mu\nu}dx^\mu dx^\nu,\nonumber
\ee
in the Einstein frame, the {\it covariant} metric tensor in the Jordan frame will now be 
of $[m]^{-2}$, and $[g^{\mu\nu}]=[m]^{2}$. This forces the redefinitions,
\be
\varphi=\alpha\exp(-\chi)\quad{\rm and}\quad\psi=\alpha\exp(-\theta),\label{Redef}
\ee
with $\alpha$ being a constant of $[m]^1$. This further requires $\chi$ and $\theta$ to be 
dimensionless, and the prior can no more be the physical Nambu-Goldstone field of the standard
St\"uckelberg theory. This also alters the N\"other current in Eq. (\ref{NC01}) to,
\be
J_\mu=\sqrt{-g}\left\{F_{\mu\alpha}g^{\alpha\beta}\partial_\beta\theta+\alpha^2\exp(-2\chi)m^2\theta\left(a_\mu+\partial_\mu\chi\right)\right\},\label{NC01}
\ee
which is still of $[m]^3$, as required physically. 
\paragraph*{}The physical difference between the previous and the present choices of field
dimensions is that in the latter case, the physical scalar field is $\varphi$, rather than
$\chi$, though the overall symmetry is intact. Thus, it can be viewed as if the roles are
shifted, and the corresponding mass generation has indirectly been shifted to the metric. 
Although a dimensional scaling of the metric can be of deeper physical interpretation, it
may find place in some special cosmological models, wherein the dual theory will be
applicable.

\section*{Conclusions}
In conclusion, it is shown that a massive Proca theory, in presence of gravity, can be re-cast as a modified
St\"uckelberg theory, coupled with STG. This composite theory is both Weyl, as well as gauge-invariant, provided
the transformation parameters are identified. This massive theory is invariant under generalized Weyl transformation, with necessary gravitational 
coupling, corresponding to a non-vanishing Weyl current. Further, the role of gravitational coupling in 
restoring this extended Weyl symmetry in otherwise a Weyl non-invariant theory, is explicated. The case of broken scale symmetry,
as a global limit of the Weyl symmetry,
is found to lead to simultaneous spontaneous breaking of gauge invariance.\\

\noindent{\bf Acknowledgements:}
We thank Dr. Vivek M. Vyas for his useful discussions during the preparation of manuscript.
KA is grateful to Prof. Samir K. Paul and Prof. Patrick Das Gupta for valuable inputs.

\end{document}